\documentclass[12pt]{article}
\usepackage[margin=.8in]{geometry}
\usepackage{authblk} 
\usepackage{comment} 
\usepackage{fullpage} 
\usepackage{xcolor}
\usepackage{subfiles}

\usepackage[font={small}]{caption}

\usepackage{xurl} 
\usepackage{hyperref}
\hypersetup{breaklinks=true, colorlinks=true, urlcolor=blue,citecolor=purple} 
\usepackage{times}
\usepackage{mathtools}
\usepackage{amsmath}
\usepackage{bm}
\usepackage{amssymb}
\usepackage{pifont}
\usepackage{marginnote}
\usepackage{graphicx}
\usepackage{dcolumn}%
\usepackage{dcolumn}
\usepackage{bm}
\usepackage{mathtools}
\usepackage{color}
\usepackage[linesnumbered,ruled,vlined]{algorithm2e}
\usepackage[utf8]{inputenc}

\usepackage[font=small,labelfont=bf]{caption}

\makeatletter
\renewcommand{\maketitle}{\bgroup\setlength{\parindent}{0pt}
\begin{flushleft}
 \fontsize{16}{32}
 \textbf{\@title}
 
   \fontsize{12}{24}
  \@author
\end{flushleft}\egroup
}
\makeatother
\newfam\bboardfam
\font\tenbboard=msbm10  
 \font\sevenbboard=msbm7
   \font\fivebboard=msbm5 
\textfont\bboardfam=\tenbboard
\scriptfont\bboardfam=\sevenbboard
\scriptscriptfont\bboardfam=\fivebboard

\begin{document}

\title{Cell density controls signal propagation waves in a multicellular synthetic gene circuit}


\author[1,*]{Marco Santorelli}
\author[2,3,*]{Pranav Bhamidipati}
\author[1] {Andriu Kavanagh}
\author[1] {Victoria Fitts}
\author[1] {Trusha Sondkar}
\author[2,4,5]{Matt Thomson}
\author[1,6]{Leonardo Morsut}

\affil[1]{Eli and Edythe Broad CIRM Center for Regenerative Medicine and Stem Cell Research, University of Southern California, Los Angles, CA, USA.}

\affil[2]{Division of Biology and Biological Engineering, California Institute of Technology, Pasadena, California, USA.}

\affil[3]{Keck School of Medicine, University of Southern California, Los Angeles, California, USA.}

\affil[4]{Department of Computing and Mathematical Sciences, California Institute of Technology, Pasadena, California, USA.}

\affil[5]{Beckman Center for Single-cell Profiling and Engineering, Pasadena, California, USA. }

\affil[6]{correspondence to: leonardo.morsut@med.usc.edu}

\affil[*]{These authors contributed equally.\vspace{0cm}}

\date{July 2021}

\maketitle
\section*{Abstract}

During organismal development, biochemical reaction networks sense and respond to mechanical forces to coordinate embryonic patterning with embryo morphogenesis. Factors such as cortical tension, cell density, and matrix mechanical properties influence differentiation and cell fate decisions by modulating gene regulatory signaling networks. A major goal in synthetic development is to construct gene regulatory circuits that program the patterning and morphogenesis of synthetic multicellular structures. However, in the synthetic context, little is known regarding how the physical properties of the growth environment impact the behavior of synthetic gene circuits. Here, we exploit physical-chemical coupling observed in a synthetic patterning circuit in order to control the size and spatial distribution of patterned synthetic cell sheets. We show that cell density attenuates the propagation of signal between neighboring cells in a multicellular sheet containing a contact-dependent patterning circuit based on the synNotch signaling system. Density-dependent attenuation leads to a signal propagation wave that exhibits distinct qualitative phases of persistent propagation, transient propagation, and no propagation. Through computational modeling, we demonstrate that cell growth parameters determine the phase of propagation observed within a growing cell sheet. Using growth-modulating drugs and spatial density gradients, we control the size of synNotch-activated cell populations and generate tissue-scale activation gradients and kinematic waves. Our study reveals that density-dependent synNotch activity can be exploited to control a synthetic multicellular patterning circuit. More broadly, we show that synthetic gene circuits can be critically impacted by their physical context, providing an alternate means for programming circuit behavior.

\section*{Introduction}
Developing tissues are fundamentally shaped by both biochemical information and the physical forces generated by tissues and the embryonic environment. Cells communicate and process biochemical signals generated by morphogen gradients through gene-regulatory networks.  Physically, cells undergoing morphogenesis receive constant physical stimuli from extracellular substrates, neighboring cells, and internal mechanics. These cellular mechanical stimuli interact reciprocally with morphogen signaling circuits to execute successful tissue patterning, which we define as the spatial organization of cells and cell fates \cite{heller_tissue_2015}. Latent pools of transforming growth factor (TGF)-$\beta$ are known to reside embedded in the extracellular matrix (ECM) and can be mobilized by high contractile tension between cells and the ECM mediated by integrins \cite{buscemi_single-molecule_2011, hinz_extracellular_2015}. TGF-$\beta$ signaling, in turn, provides input to pathways controlling cell motility and cortical tension such as Rho GTPases, PAK, and PI3K/AKT/mTOR \cite{lamouille_molecular_2014}, and it can also cause RhoA-mediated tight junction disassembly and increased epithelial cell plasticity through its interaction with Par6\cite{ozdamar_regulation_2005}. These reciprocal influences contribute to cell state and constitute mechanical-chemical and chemical-mechanical coupling, respectively.

Mechanical-chemical coupling is indeed at our very fingertips. Vertebrate digit numbering is sometimes described as an example of chemical morphogen patterning. Digit precursors are thought to develop due to a reaction-diffusion network of morphogens WNT, BMP, and SOX9 that interact to form a spatially periodic Turing pattern fanning across the vertebrate limb bud, the crest of each wave become a digit \cite{sheth_hox_2012, raspopovic_digit_2014}. Although morphogens are the effectors, the number of resulting digits is roughly the width of the limb bud divided by the characteristic wavelength of the Turing system. Therefore, vertebrate digit numbering can also be considered  an example of mechanical-chemical coupling due to the crucial boundary conditions supplied by limb bud geometry. Bulk cell density, \cite{kempf_bulk_2016, tariki_yes-associated_2014} applied mechanical stresses \cite{kinoshita_mechanical_2020, cohen_mechanical_2020}, boundary geometry \cite{etoc_balance_2016}, and topological defects \cite{saw_topological_2017} have also been shown to influence cell fate.

Do synthetic morphogenetic circuits exhibit similar sensitivity to physical context? In recent years, synthetic patterning circuits have been devised to pattern multicellular cell populations \cite{morsut_engineering_2016,sekine_synthetic_2018, toda_programming_2018, toda_engineering_2020}. Many such systems use synNotch, an engineered signaling system, to build self-organizing synthetic cellular structures. SynNotch was developed from endogenous Notch, a single transmembrane domain receptor that reponds only to membrane- or ECM-bound ligands due to its mechanosensitive properties \cite{morsut_engineering_2016, stassen_notch_2020}. Because synNotch retains the contact-dependent nature of Notch, it stands to reason that its activity may be responsive to conditions that affect cell mechanics, as are other mechanoreceptors such as cadherins\cite{yap_mechanosensing_2018}. However, the effects of ambient physical factors on synNotch activity remain poorly understood. 

Here, we investigate the impact of the physical culture environment on synNotch activity in a multicellular signaling context. We show that cell density attenuates the propagation of signal between neighboring cells through a contact-dependent synNotch patterning circuit. Density-dependent attenuation leads to a signal propagation wave that exhibits distinct qualitative phases of persistent propagation, transient propagation, and no propagation. Through computational modeling, we demonstrate that cell growth parameters determine the phase of propagation observed within a growing \textit{in vitro} cell sheet. Using growth-modulating drugs and spatial density gradients, we control the size of Notch-activated cell populations and generate tissue-scale activation gradients and kinematic waves. Our study reveals that density-dependent synNotch activity can be exploited to control a synthetic multicellular patterning circuit. More broadly, we show that synthetic gene circuits can be critically impacted by physical context, providing an alternate means for programming circuit behavior.

\section*{Results}


\subsubsection*{Cell density attenuates signal transduction by SynNotch}

To interrogate the relationship between physical aspects of growth conditions an the activity of synNotch-based synthetic morphogenetic circuits, we developed an assay for synNotch activation in murine fibroblasts based on a previously reported Sender-Receiver signaling paradigm \cite{morsut_engineering_2016}.  Briefly, Sender cells constitutively expressing membrane-bound GFP activate Receiver cells containing a synNotch receptor with a GFP-binding domain in place of the Notch extracellular domain (NECD) (Figure 1A, left). Upon contact-dependent activation, the intracellular portion of the receptor (the transcription factor tTA) translocates to the nucleus and activates expression of cytosolic mCherry, providing a readout of synNotch receptor activity. To study the effects of different physical inputs, a screen was performed by co-culturing Senders and Receivers in a 1:1 ratio while changing total cell density, extracellular matrix (ECM) stiffness, ECM composition (PDMS, Fibronectin, Gelatin, or Matrigel), or cytoskeletal tension via addition of ROCK-inhibitor, blebbistatin, or Latrunculin-A. mCherry intensity in Receivers was then quantified by fluorescence-activated cell sorting (FACS) at 24hrs (Figure 1A, right). 

We found that cell density, but not other physical variables, strongly attenuated synNotch activation. Specifically, the median mCherry fluorescence drops as cell density increases above $1.2\text{ x }10^3\text{ cells/mm}^2$ and is fully suppressed above $2.4\text{ x }10^3\text{ cells/mm}^2$ (Figure 1B; black dots indicate median). We also observed bimodal fluorescence distributions at and below $2.4\text{ x }10^3\text{ cells/mm}^2$, representing subpopulations of activated and quiescent Receivers, while the activated subpopulation is largely absent from distributions at higher densities. In contrast, stiffness and composition of ECM did not have appreciable effects on Receiver activation (Figure 2C-D). ROCK-inhibitor and blebbistatin did not have appreciable effects on synNotch activation, and Latrunculin-A moderately decreased Receiver activation. A confluent fibroblast culture was found to have a density of $1.25\text{ x }10^3\text{ cells/mm}^2$ (see Supplementary Figures), indicating that synNotch activation is inhibited at super-confluent cell densities. Hereafter, we report cell densities in relation to the density of a confluent fibroblast cell sheet; for example, "2x confluence" will indicate a density of $2.5\text{ x }10^3\text{ cells/mm}^2$.

\subsubsection*{Mathematical modeling of a multicellular signal propagation circuit with density-dependence}

Next, we developed a mathematical model to ask how density-dependent attenuation of signaling might impact the propagation of synNotch signaling through a multicellular circuit. We devised a mathematical model of a putative "Transceiver" cell that can act as both Receiver and Sender of a signal ligand (Figure 2A). In our model, Transceiver activation by the signal activates transcription of the signal, which is produced after a time delay. The signal then acts on neighboring cells, producing a signaling wave radiating from an initiating Sender cell. The effect of density is modeled as an exponential attenuation of signal perception with density. Mathematically, we model the cells as a system of delayed differential equations (DDEs) where we represent ligand concentration in cell "$i$" as $S_i$, and the production rate of $S_i$ evolves in each cell according to a DDE.

\begin{align}
\frac{\mathrm{d}S_i}{\mathrm{d}t} &=  f\left(t-\tau\right) - S_i
\end{align}

In this dimensionless equation, the signaling ligand is produced at a rate determined by the activation term $f$ after a time delay $\tau$ and is subject to degradation in the term $-S_i$. The activation term 

\begin{align}
f(t) &= \lambda + \alpha \cdot \frac{\left(e^{-\rho} \, I_i\right)^p}{k^p + \left(\delta\, S_i\right)^p + \left(e^{-\rho}\, I_i\right)^p}
\end{align}

is sigmoidal in $I_i$, a weighted mean of the signal input to cell "$i$" from its neighbors. $k$ and $p$ are respectively the threshold and cooperativity of the sigmoid, and $\lambda$ and $\alpha$ are basal and inducible production rates. Furthermore, $f$ includes an inhibitory term $\delta S_i$ through which the signaling ligand suppresses its own signaling receptor, a phenomenon reported in the literature as \textit{cis}-inhibition \cite{sprinzak_cis-interactions_2010}. In our model, synNotch signaling is exponentially inhibited by density through the term $e^{-\rho}$, where $\rho = \left(\text{cell density}\right) / \left(\text{cell density at confluence}\right)$. 

Given the cell density, we create a 2D lattice of cells and calculate cell-cell contacts based on an interaction radius $r_\text{int}$. Starting from random initial conditions (Supplemental Figures), we integrate the ODEs over time using the method of steps and numerical integration using the Euler method. We then examine circuit activation by calculating the area of cells that exceed a threshold $S_i > k$. Simulation of $n=2500$ cells ($1$ Sender and the remaining Transceivers; $r_\text{int}=3$) was performed in this fashion at cell densities of 1x, 2x, and 4x confluence up to simulation time $t=3$.

The simulation renderings shown in Figure 2B show that, when induced by a single Sender (magenta), the signal ligand GFP indeed propagates outward over time in a wave that is inhibited at super-confluent densities. Figure 2E shows a plot of signal activation area as a function of time, and the continual rise in the curve labeled "1x confluence" indicates the area of the activated cell region increases over time. At time $t=0$, the Sender begins expressing a constant amount of $S$, and Transceivers begin to respond after a delay $\tau$. The other curves show area over time at 2x and 4x confluence, which produce slower and no wave propagation, respectively. These results suggest that density could be used as an additional factor to control the area of Transceiver patterning without interfacing with the circuit directly. 

\subsubsection*{An \textit{in vitro} Transceiver patterning circuit using synNotch shows cell density-dependent activation}

To verify our observations experimentally, we implemented the Transceiver patterning circuit in fibroblasts and tested its response to variation in cell density. In order to generate the transceiver cell line, we stably integrated three transgenes. The first one constitutively expresses a synNotch receptor with an anti-GFP nanobody as the extracellular domain and the transcription factor tetracycline transactivator (tTA) as the intracellular domain. The second transgene expresses membrane bound GFP (synNotch cognate ligand) under control of the TRE promoter, which is activated by tTA. Finally, the third transgene expresses TRE-driven cytosolic mCherry as a readout of synNotch activity and constitutively expressed tagBFP. When it comes into contact with GFP on the membrane of another cell, a Transceiver should begin to produce GFP in turn, causing a chain reaction (Figure 3C). Transceiver function was evaluated by co-culturing with Sender cells in a 1:1000 ratio (Sender:Transceiver) and performing high-magnification time-lapse imaging centered on individual Sender cells. The area of GFP fluorescence was then calculated using an image analysis workflow (see Methods and Materials). 

Fluorescent imaging revealed that when stimulated by a Sender cell, Transceivers indeed produce GFP in a propagating wave that travels outward in the sheet (Figure 2D). When total cell density is increased to $2$x confluence ($2.5\text{ x }10^3\text{ cells/mm}^2$) and $4$x confluence ($5\text{ x }10^3\text{ cells/mm}^2$), we observe that the propagation waves are respectively slowed and entirely inhibited (Figure 2D). In Figure 2F, similar to computational simulations in Figure 2E, the area of activation drops considerably at $2$x confluence and is virtually absent at $4$x confluence, demonstrating that Transceiver activation waves can be triggered and controlled \textit{in vitro}.

\subsubsection*{Cell population growth attenuates multicellular signaling propagation on long time-scales}

Over the course of experiments, we observed an increase in cell density over time due to cell growth as well as a long time-scale attenuation of the multi-cellular signaling wave (Figure 3). To investigate the relationship between cell growth and signal attenuation, we co-cultured Senders and Transceivers (1:1000 ratio) at an initial density of 1x confluence. We performed time-lapse imaging over many days of isolated propagating waves using brightfield capture and epifluorescence imaging of GFP, mCherry, and miRFP (expressed by Senders). In Figure 3A we show one propagating wave over many days and find that Transceiver GFP signal starts to decline at day 4-5 of growth in culture (Figure 3A, "GFP"). Interestingly, while GFP expression decays almost fully by the end of the time-course, downstream mCherry reporter expression (Figure 3A, "Reporter") remains in the area once occupied by activated Transceivers. An image analysis pipeline was then used to construct a mask of activated GFP, shown in Figure 3A, "Density" as a black mask superimposed on brightfield images of the cell sheet. In Figure 3C, this procedure was repeated for a total of 5 replicates, and distributions of activated area are plotted as a function of time to demonstrate the full time-course of propagation and attenuation of ligand expression.
 
To study how cell density dynamics affect circuit behavior in principle, we analyzed cell growth through the lens of logistic growth, a classic paradigm of population growth used in population dynamics and ecology. In Figure 3B, Transceivers were cultured at an initial density of 1x confluence, and cell density was measured daily using an automated cell counter. Growth data were used to parameterize the logistic growth equation (Figure 3B; schematic of logistic growth in Figure 4B), which could then be plugged into the model. Simulation time was then converted to days under an assumption that the dilution/degradation rate of signal protein can be approximated by the growth rate. To represent activation of mCherry downstream of synNotch, each cell was modeled with an additional Reporter species $R_i$ produced in response to $S_i$ according to an ordinary differential equation (ODE).

\begin{align}
\frac{\mathrm{d}R_i}{\mathrm{d}t} &= \gamma_R\,(S_i - R_i)
\end{align}

Here, the speed of response is set by $\gamma_R$. Observing that mCherry has slow kinetics, we set $\gamma_R=0.1$. With these changes, the model was then initialized as in Figure 2B with $n=1225$ cells ($1$ Sender and the remaining Transceivers) and simulated over a longer time-course, up to 7 days. 

In the model, as in experiments, we found that population growth leads to attenuation of the Transceiver signaling wave and decay of signal expression. The area of cells expressing GFP was calculated as in Figure 2E, sub-sampled daily, and plotted in Figure 3C over the experimental observations. Figure 3D, "GFP" shows renderings of the model at daily time-points that exhibit the same properties as the experiment in 3A, initial propagation of GFP signal followed by attenuation. Reporter expression (Figure 3D, "Reporter") also corresponds closely to the \textit{in vitro} phenotype, persisting after the GFP signal decays until the end of the time-course. The onset of attenuation is slightly anticipated in the simulation relative to the experiment, occurring on day 3. Overall, the effect of cell growth, modeled as a logistic process, was sufficient to recapitulate the transient propagation observed in experiments.

\subsubsection*{\textit{In silico} exploration of growth parameters reveals distinct phases of activation explained by a critical density}

In order to more specifically dissect how growth over time can affect the qualitative behavior of synthetic cell sheets in theory, we generated a "phase diagram" of circuit behavior across $600$ combinations of initial cell density ($\rho_0$) and growth rate ($g$). For each of parameter set, we simulated $5$ replicates with different random initial conditions up to a time of $t=\text{6.9 days}$ ($n=\text{10,000 cells}$, $1\%$ randomly chosen Senders, $r_\text{int}=1$). The third parameter of logistic growth (the carrying capacity) was set to $\rho_\text{max} = 5.63\text{x confluence}$ based on the logistic model fit (Figure 3B). We then classified dynamics into distinct qualitative phases by studying Transceiver behavior at early and late time-points. Initial activation of a lattice was calculated by applying a threshold to the mean change in $S$ ($dS/dt$) at $t = \tau$, the first time-point of activation, and attenuation over time was calculated by applying a threshold to mean $S$ expression at the end of simulation time (see Supplementary Information for the thresholding procedure). Parameter sets showing initial activation were classified as "Persistent Propagation" if they did not eventually attenuate and "Transient Propagation" if they did. All parameter sets showing no initial activation were also attenuated at the end of simulation time, and these were classified "No Propagation." To quantify Transceiver activation over time, we also calculated the average percentage of Transceivers with $S$ expression greater than the signaling threshold $k$.

Each region of the resulting phase diagram demonstrates a major phase of circuit behavior. Circuits in the dark blue region in Figure 4A exhibit persistent propagation, in which waves do not stop propagating and eventually saturate the lattice. Circuits in the blue region, like the experimental system, generate propagation waves that eventually attenuate. The gray region represents circuits that failed to activate. Examples from these phases are shown in Figure 3B. We observe that no activation occurs above a certain critical density of the lattice ($\rho_\text{crit}\approx 3.5\text{x confluence}$). Equipped with the growth parameters of the logistic equation (Fig 4C), the phases can be understood conceptually by simply considering where the density growth curve lies relative to $\rho_\text{crit}$. As illustrated in Figure 4D, when $\rho_\text{crit} < \rho_0$ (gray curve), density always exceeds the critical value and no propagation occurs, resulting in a horizontal phase boundary at $\rho_0=\rho_\text{crit}$. Below this boundary, the phase is determined by whether or not the population reaches $\rho_\text{crit}$ (at a time $t_\text{crit}$) by the end of the experiment. At other values of $\rho_\text{max}$, $\rho_\text{crit}$ remains unchanged, while the boundary between transient and persistent propagation is shifted (Supplementary Figures).

\subsubsection*{Initial cell density modulates area of Transceiver-activated cell population}

Equipped with these suggested control parameters from the model, we tested whether the area of GFP expression can be controlled by plating Transceivers at different initial cell densities. Time-lapse imaging and quantitative measurements of propagation area were performed as in Figures 3A and 3C, and computational simulations were performed as in Figures 3C and 3D, now at initial densities of $2\text{x confluence}$ and $4\text{x confluence}$. The resulting area of activation for both modalities were quantified as described above. Representative experimental time-courses are shown in Figure 5A (Senders in magenta and transceiver GFP ligand in green). Attenuation can be seen in all three conditions, but the quantitative analysis in Fig 5C shows that it occurs earlier in the two higher density conditions. We therefore discovered that the area of ligand activation can indeed be modulated over time simply by changing the initial cell density. 

\subsubsection*{Growth rate-modulating drugs affect the qualitative phase of Transceiver activation dynamics}

The computational model also predicts that cell growth rate modulates Transceiver activation phase. Experimentally, we cultured Senders and Transceivers (1:100 ratio) in the presence of ROCK-inhibitor (Ri) and fibroblast growth factor 2 (FGF2), which suppress and stimulate fibroblast growth, respectively, and were shown by FACS analysis to have minimal effect on synNotch activation (Figure 1 and Supplementary Figures). Under each condition (untreated, Ri, and FGF2), cell density was counted daily and the total area of GFP expression was calculated as described above for Figures 3B and 3C. Using the cell density measurements, logistic regression was performed to produce empirical growth parameters as described above for Figure 3B. These were then used to generate a theoretical simulation of Transceiver signal propagation ($n$ = \text{22,500}, $1\%$ Senders (randomly chosen), $r_\text{int}=1$), for comparison with experiment.


We found that these conditions generated a range of cell growth rates (Ri $<$ untreated $<$ FGF2) that recapitulated the three propagation phases found by \textit{in silico} analysis. The growth curves and model fits in Figure 5D show that FGF2 (green curves) induces faster population growth that reaches a moderately lower carrying capacity. In contrast, Ri treatment (purple curves) leads to pronounced suppression of growth. Observing the propagation area in Figure 5E, we see FGF2 causes Transceivers to reach signaling attenuation sooner, while Ri-treated Transceivers exhibit persistent propagation over the entire time-course. These observations are consistent with the conceptual model of a critical density. \textit{In silico} propagation area over time under drug conditions (Figure 5F) exhibits similar behavior. Ri treatment caused persistent, saturating propagation, while FGF2 treatment led to earlier signaling attenuation. We note that experimental propagation in the presence of FGF2 was more suppressed than in the computational model, suggesting that FGF2 may have additional effects on signaling that were not accounted for in the model. Overall, we find that these results support the hypothesis that modulation of cell density through growth-modulating drugs can control propagation behavior of Transceivers. 

\subsubsection*{Tissue-scale gradients of cell density produce long-range activation gradients and kinematic waves}

Finally, we explored whether density-dependent signaling waves could be used to generate global patterns at the scale of a developing tissue. Different volumes of Sender-Transceiver 1:100 mixtures were plated onto culture wells either normally or preferentially on one side. The resulting co-cultures exhibited whole-well gradients of cell density, with mean initial densities of $1\text{x confluence}$ and $2\text{x confluence}$. Wells were then imaged using high-magnification brightfield capture and fluorescence imaging in GFP and BFP channels. GFP fluorescence along the gradient was studied by image analysis, while constitutively produced BFP was used as a readout of cell density.

We observed that a spatial gradient of cell density produces long-range activation gradients and kinematic waves. In Figure 6A, culture wells with an average initial density of $2\text{x confluence}$ are shown with flat (Row 1) and graded (Row 2) density distributions, which can be observed in the BFP channel (blue indicates low BFP fluorescence). While the flatly-distributed well shows relatively uniform Transceiver activation throughout the well, the gradient-distributed well shows preferential activation in the top region, where density is lower. Thus, we observe a gradient of activation spanning the diameter of the well. Figure 6B shows multiple imaging time-points of a graded-density well with an average initial density of $1\text{x confluence}$. Signal propagation begins at the very top of the well, where density is highest, as evidenced by the BFP channel, and spreads to the bottom of the well over the course of days 1-5. This phenomenon is quantified in Figure 6C, where each colored curve represents the profile of GFP fluorescence along the gradient axis at a specific time. The low-density areas of the well likely fail to generate signal propagation due to the necessity for cell-cell contact with a Sender, which becomes rare as cells become more dilute (see Supplementary Figures). In principle, these results reflect a likely range of optimal density that lies between excess cell dilution and high-density inhibition (Figure 6D). As cell growth occurs throughout the well, more dilute regions near the bottom enter the optimal range as more dense regions at the top exit. This signaling wave likely only appears to have long-range coordination but is in fact a cell-autonomous "kinematic" wave.

\section*{Discussion}

In this study, we investigated the impact of physical properties on the propagation of signaling waves through a synthetic signaling circuit. We found that signal propagation through a synthetic cell population is determined by cell density. By manipulating cell density through external signals and spatial patterning, we are able to modulate the architecture of the signaling wave in space and time. We observe distinct phases of propagation that we call persistent propagation, transient propagation, and no propagation. Spatially, we achieve long-range signal gradients and kinematic waves on the scale of an embryo. Our study raises the key question of how physics impacts synthetic gene circuits. In future work it will be important to understand how a broader array of physical properties can impact circuit behavior and similarly how circuits can be applied to sculpt the physical properties of tissues. Dissecting how mechanical-chemical coupling operates on synthetic circuitry such as synNotch will no doubt also be important for synthetic biology more broadly, as it enters a new phase of industrial and biomedical applications.

\begin{figure}
    \centering
    \includegraphics[width=0.6\textwidth]{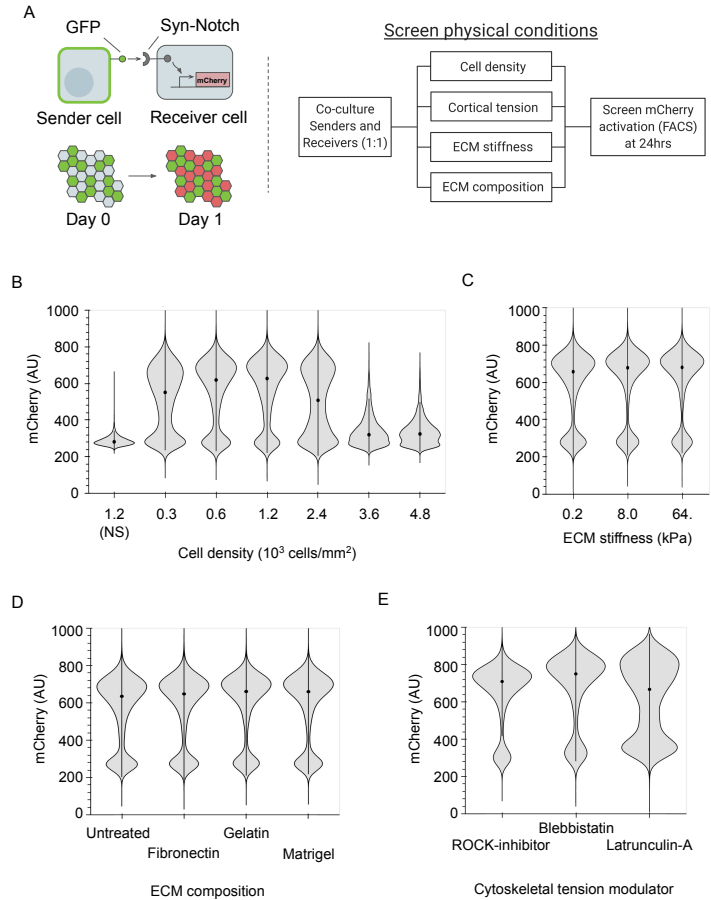}
    \caption{\textbf{SynNotch activity is inhibited by cell density and unaffected by growth substrate or cytoskeletal tension modulators.} (A) Conceptual scheme of Sender-Receiver synNotch signaling. In synNotch receptors, both the extracellular and intracellular domains of the contact-dependent receptor Notch can be exchanged for differential receptor and transcription factor domains, respectively. Left: Sender cells constitutively present membrane bound GFP on the extracellular membrane surface. In Receiver cells, the transgene anti-GFP-synNotch-tTA is constitutively expressed. Upon binding GFP at its receptor domain, synNotch undergoes proteolysis. The freed intracellular domain translocates to the nucleus and activates the target gene(s) (cytosolic mCherry in the case of Receivers). Together, Senders and Receivers provide a readout of synNotch activity. Right: A screen was performed to evaluate the sensitivity of synNotch to physical perturbations. Senders and Receivers were plated in a 1:1 ratio and cultured under different conditions for 24hrs. mCherry fluorescence was then measured by FACS to quantify synNotch activity. (B-E) Violin plots of mCherry fluorescence. Black dot indicates median fluorescence. In each graph, a different external or internal physical condition is assayed. (B) Bulk cell density: synNotch activation is reduced above an optimal density of $1.2\text{ x }10^3\text{ cells/mm}^2$ and fully suppressed at densities above $2.4\text{ x }10^3\text{ cells/mm}^2$.  NS = no Senders (only Receivers). (C-D) PDMS stiffness and other ECM compositions: no appreciable effect on synNotch activation. (E) Cytoskeletal tension modulators: ROCK-inhibitor (100 uM) and blebbistatin (25 ug/ml) showed no appreciable effect on synNotch activation. In the presence of Latrunculin-A (200 uM), basal fluorescence is moderately increased and Sender-induced fluorescence is slightly reduced.}
    \label{fig:1}
\end{figure}

\begin{figure}
    \centering
    \includegraphics[width=0.6\textwidth]{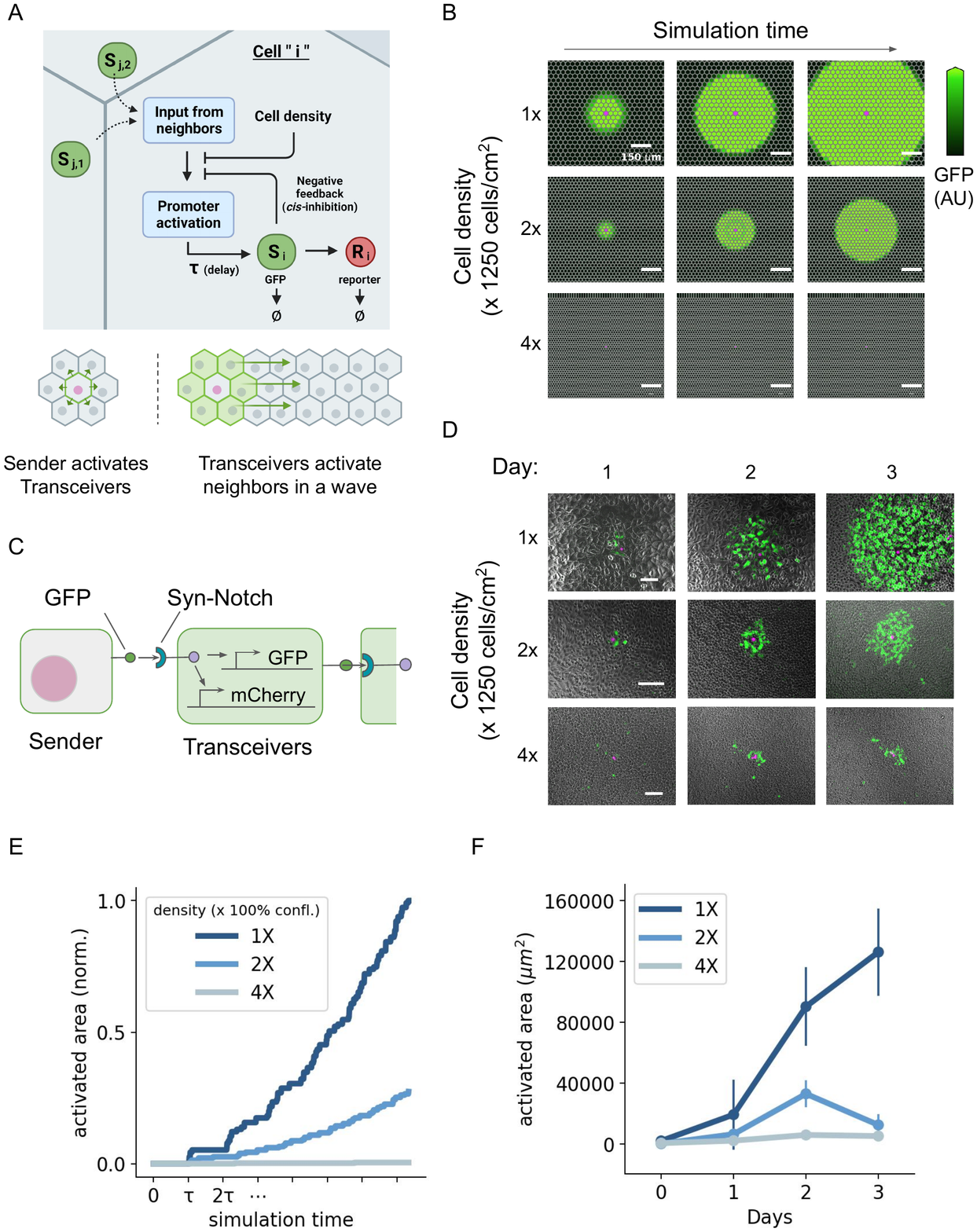}
    \caption{\textbf{Density-dependent activation of a multicellular signal propagation circuit using synNotch. } (A) Mathematical model of a “Transceiver”. Above: A dynamical model was developed to understand the effect of density-dependence on a juxtacrine circuit with both Sender and Receiver capabilities (“Transceiver” signaling circuit). Cells were modeled as regions on a 2D hexagonal mesh producing a signaling molecule “S”, which can be perceived up to $r_\text{int}$ neighbors away, and a reporter molecule $R$. At each time-step, a given cell $i$ with an amount of signaling molecule $S_i$ experiences production and degradation-dilution of $S$ and $R$. $S$ production rate is a non-linear increasing function of $S$ expression in cell $i$’s neighbors, and $R$ production is a linear function of $S$. Both $S$ and $R$ degradation-dilution rates follow linear, first-order kinetics. While all other reactions are immediate, S production requires a time interval $\tau$ for transcription and translation to occur, making this a delayed differential equation (DDE) model. The production rate of $S_i$ is also inhibited by $S_i$ itself, mimicking the negative-feedback effect seen in Notch-ligand systems known as cis-inhibition. The amount of $S$ perceived by a cell is also suppressed by the ambient cell density $\rho$. Below: Transceiver cells generate traveling waves of activation in a multicellular context. When activated by a cell constitutively expressing $S$ (“Sender” cell), Transceiver cells produce high amounts of $S$. In a tissue consisting of Transceivers, this triggers a traveling activation wave. (B) Rendering of mathematical model at constant cell densities. Renderings show Transceiver activation indicated by simulated GFP intensity. Each frame indicates a single simulation time point ($t$ = $1$, $2$, $3$). Transceivers are activated by a single Sender cell (magenta) with constant S. Simulations performed at indicated densities. As density increases, the speed of the activating wavefront decreases, and ultimately activation does not occur at all. (C) Schematic of experiment, On the left, the sender cell marked with a purple nucleus and constitutively expressing the GFP ligand; on the left, two transceiver cells marked in green, constitutively expressing synNotch receptor driving the expression of its own ligand and an mCherry reporter. (E) Quantification of \textit{in silico} signal propagation in terms of area of activated cells and corresponding experimental data (F). (n=5 for each density). In both experiments and simulation, wave propagation decays as a function of cell-density. }
    \label{fig:2}
\end{figure}

\begin{figure}
    \centering
    \includegraphics[width=0.6\textwidth]{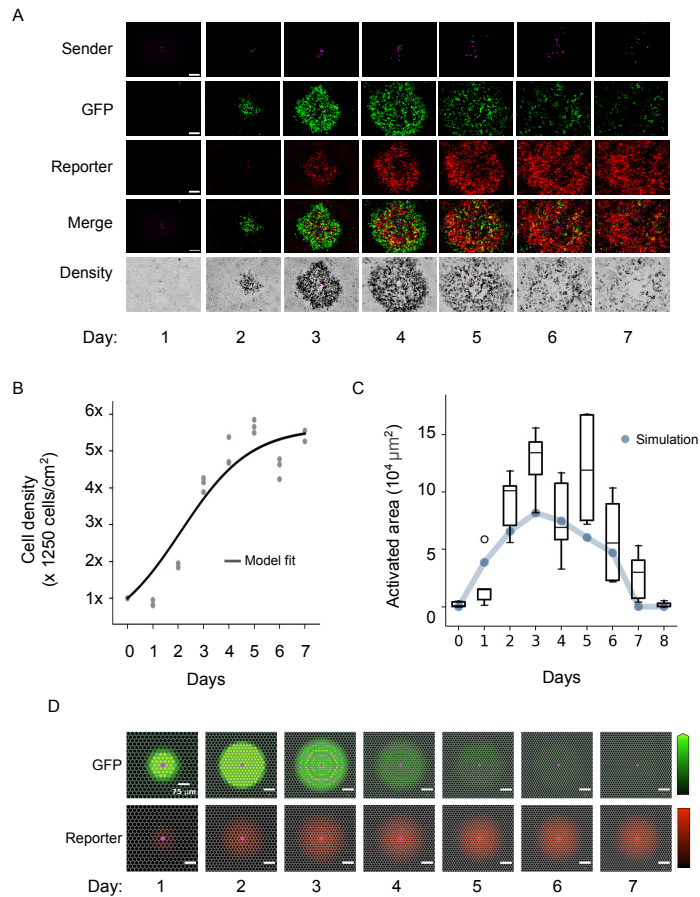}
    \caption{\textbf{Cell population growth over time leads to transient activation.} (A) Time-lapse images of SynNotch wave propagation over 7 days vs cell density. Purple indicates the sender cells nuclei, green indicates the ligand and red the cytosolic syn-Notch reporter gene. Gray signal in the bottom lane marks the cell density and how it changes over time (the gray intensity is the readout of the activated area and it has not to be intended as quantification of the cell density). Wave initially propagates but then decays over time as cell density increases (scale bar 100 um). (B) Quantification of cell density over time (gray circles). Data were used to parameterize the logistic growth equation (solid line). (C) Experimental quantification of activation area over time from images. Box-and-whisker plots indicate the distribution of activated areas of single activation spots at each time-point (hollow circles indicate outliers). Blue line shows model-simulated activation area for a single activation spot (n = 1225 cells, including one Sender). Density changes as a function of time according to the logistic growth equation parameterized in (B). Time-units of simulation were converted to days by setting the S degradation-dilution rate equal to the logistic growth rate g. (D) Rendering of model simulation over 7 days. The activation spot initially expands until density increases sufficiently to cause signal attenuation. The slower kinetics of the reporter molecule allow it to slowly accumulate and persist, keeping a record of the previously activated area. Magenta = Sender cell.
}
    \label{fig:3}
\end{figure}

\begin{figure}
    \centering
    \includegraphics[width=0.6\textwidth]{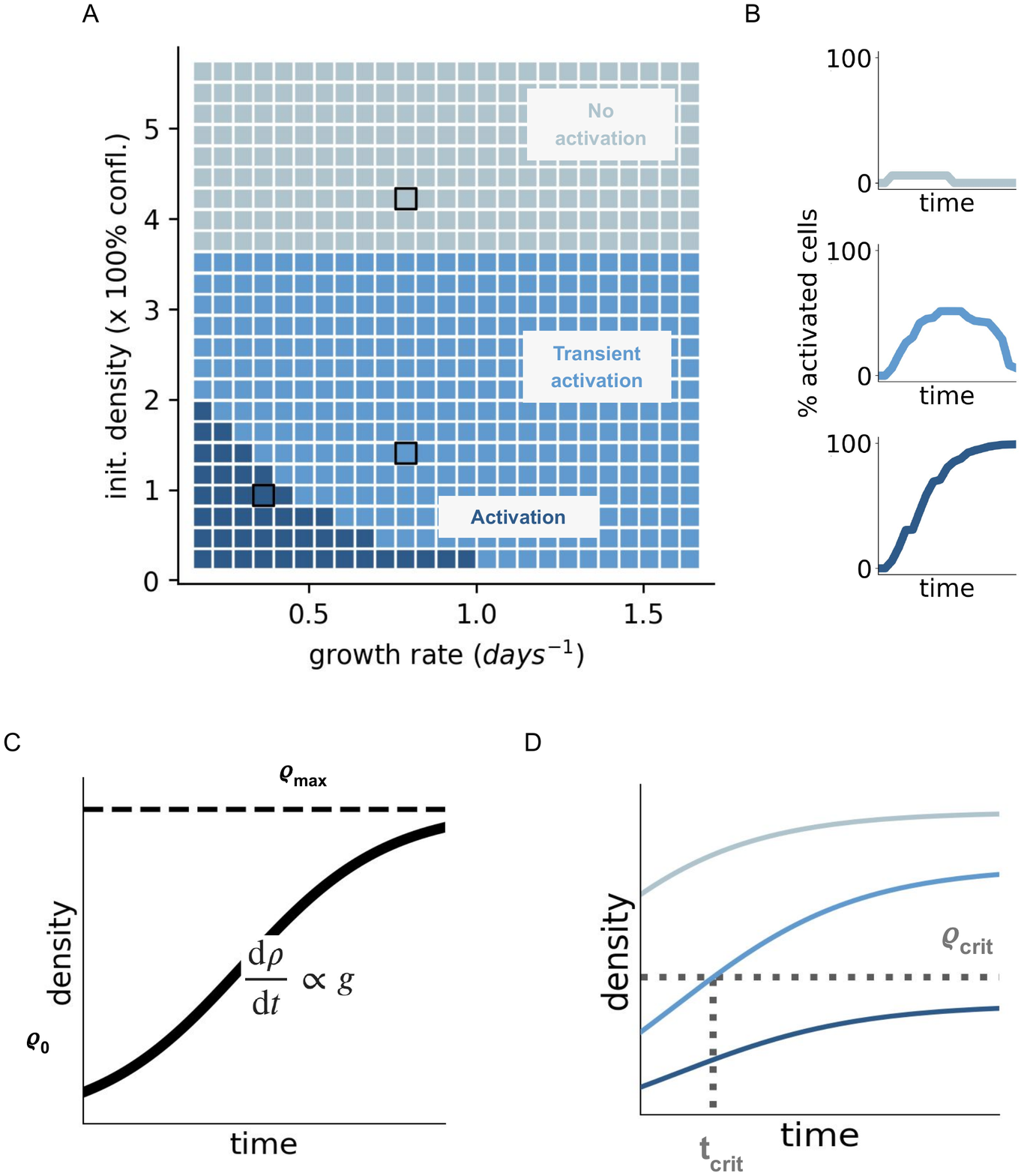}
    \caption{\textbf{Transceiver activation dynamics can be controlled \textit{in silico} by changing growth parameters.} (A) Phase diagram of Transceiver activation dynamics in a multicellular context. To investigate the effect of growth parameters on the overall activation dynamics of a transceiver tissue, a mesh of cells was simulated under changing density conditions with different values of $\rho_0$ and $g$ ($600$ parameter sets; $n$= 10,000 cells, $N$ = 5 replicates). $1\%$ of cells were randomly chosen to be Senders (the rest Transceivers), and the logistic model fit (Fig. 3B) was used to set the carrying capacity $\rho_\text{max} = 5.63$ ($7038\text{ cells/mm}^2$). After simulation, signal $S$ was averaged for each parameter set at each time-point across all Transceivers and replicates. The overall activation dynamics were classified into $3$ behaviors, or phases: unresponsive to Senders (“No activation”), an initial response followed by attenuation (“Transient activation”), and an initial response without attenuation (“Activation”). Initial activation was determined by thresholding the mean initial change in $S$ ($dS/dt$ at $t = \tau$). Attenuation was determined by thresholding the mean $S$ at the final simulation time-point ($t = 6.9 \text{ days}$). (B) Examples of propagation over time for each observed phase. (C) The parameters of the logistic growth equation determine change in density over time. A logistic growth process starts at an initial density $\rho_0$ and behaves like exponential growth with a growth rate $g$ at low population density. As population density increases, growth becomes asymptotic as $\rho$ approaches the population’s carrying capacity $\rho_max$. For a growth process, $\rho_0 < \rho_\text{max}$. The logistic equation thus approximates the dynamics of a growth process across all growth phases. (D) Transient activation is explained by a critical density $\rho_\text{crit}$ above which Transceivers can no longer activate one another. If $\rho_\text{crit} < \rho_0$ ,  cell density is always too high for activation to occur (no activation). Similarly, if $\rho_\text{crit} > \rho_\text{max}$ ,  Transceivers will not become suppressed by growth (persistent activation). However, if $\rho_0 < \rho_\text{crit} < \rho_\text{max}$, Transceiver signaling will attenuate at some time $t_\text{crit}$.}
    \label{fig:4}
\end{figure}

\begin{figure}
    \centering
    \includegraphics[width=0.6\textwidth ]{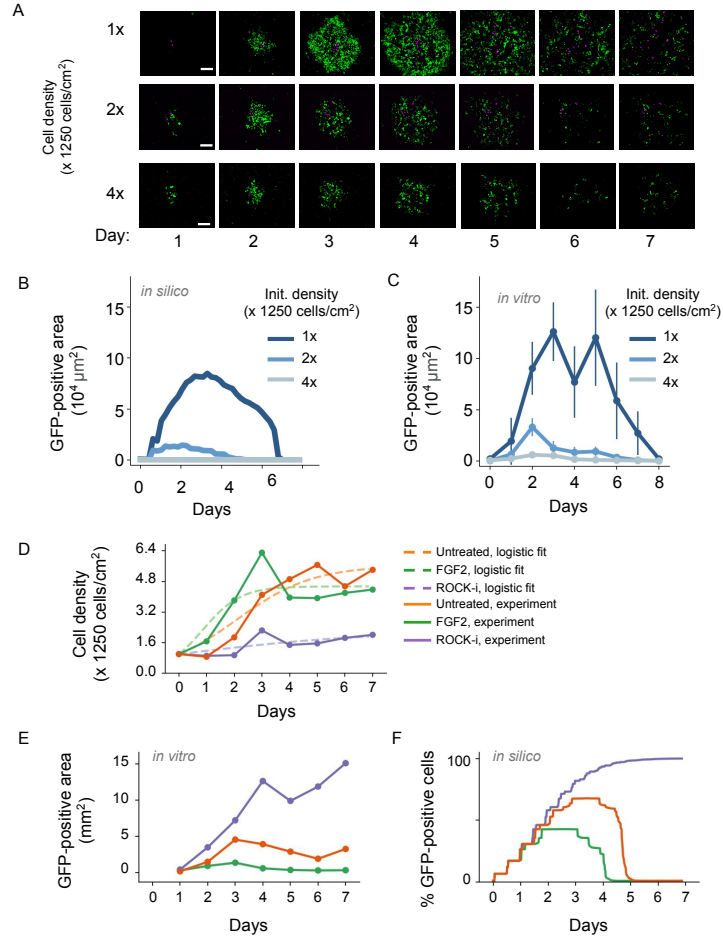}
    \caption{\textbf{\textit{In vitro} control of Transceiver activation area by manipulating population growth.} (A) Time-lapse images of a SynNotch wave propagation over $7$ days vs cell density. Purple indicates the sender cells nuclei, green indicates the synNotch cognate ligand. Waves decays faster as function of the initial cell density (scale bar 100 um). Row 1 reproduced from Fig. 3A, "GFP". (B) Model-predicted activation area of a single spot at different densities. GFP activation area shrinks with increasing plating density. Simulation was performed on a hexagonal mesh of Transceiver cells ($n = 1935$) surrounding a single Sender. GFP-positive area was calculated as the area occupied by cells with $S$ expression above the Hill promoter threshold value k. (C) Time-series analysis of mean single-spot activation area at different densities. $n=5$ replicates. Error bars indicate standard deviation. (D) Time-series analysis of cell density. Co-cultures of Senders and Transceivers (1:100) were grown with no added drugs, FGF2 (250 ng/ml), or ROCK-inhibitor (50 uM). Density at plating was $1250 cells/mm^2$, and density counting was performed daily using an automated cell counter. Data were used to parameterize the logistic growth equation (dashed lines). "Untreated" curves reproduced from Fig. 3B. (E) Time-series analysis of bulk activation area under growth-modulating drug conditions: untreated, FGF2 (250 ng/ml), or ROCK-inhibitor (50 uM). (F) Simulated effect of growth modulation on Transceiver activation. Growth parameters identified from the density time-course in (D) were used to simulate signaling behavior of a mesh of cells ($n$ = 22,500), of which $1\%$ of were randomly chosen to be Senders (the rest Transceivers). Simulation was performed up to $t = 6.9$ days ($r_\text{int} = 1$). Transceivers with signal expression greater than the circuit’s threshold ($S > k$) were labeled GFP-positive.}
    \label{fig:5}
\end{figure}

\begin{figure}
    \centering
    \includegraphics[width= .7 \textwidth]{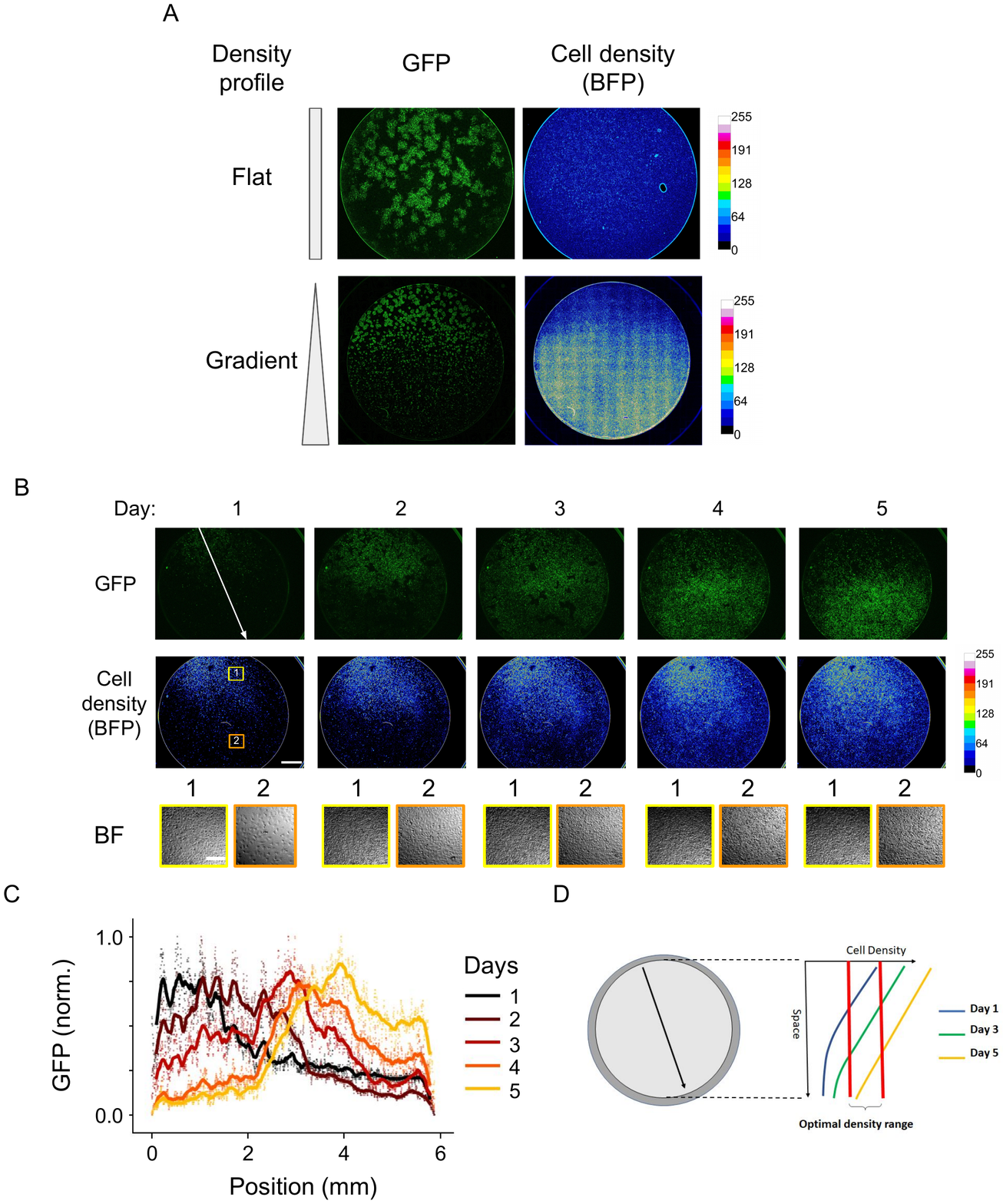}
    \caption{\textbf{Gradients of cell density produce tissue-scale activation gradients and kinematic waves.} (A) Whole well images showing “tissue” level synNotch patterning in different cell density conditions. GFP signal (left) indicates synNotch activation area; BFP signal (right) is the cell density read-out, calibration bar ranging from 0 (black) to 255 (white) indicates less and more dense areas respectively. In the control experiment (upper lane) cells were uniformly plated in the well, conversely in the lower lane cells were plated by following a density gradient. (B) Time lapse images of the whole well showing a kinematic wave guided by a cell density gradient. GFP signal (upper lane) indicating activated areas, over 5 days, travel from the upper side of the well to the bottom; BFP signal (middle lane) indicates the cell density, changing over time accordingly to the growth curve shown in figure 5D (scale bar 1 mm). The brightfield images (lower lane) show the cell density in the yellow quadrants 1 and 2 reported in the BFP lane (scale bar 200 um). (C) Plot profile measured along the white arrow reported in Fig. 6B (upper left image) over 5 days. On the Y axes are reported the normalized GFP intensity values; On the X axes the distance from the upper side to the lower side of the arrow. (D) Schematic representing the rationale behind experiment in (B). On the left is represented the well and the direction of the cell density gradient (decreasing along the arrow direction), on the right side, is represented the dynamical change of the cell density in space and time. The optimal density range (in between the two red lines) is the range of cell density where the propagation works better; below this range, cells do not touch to each other and they do not communicate, over this range cell density is to high and as shown in figure 1B, synNotch pathway is not efficiently activated.}
    \label{fig:6}
\end{figure}

\subsection*{Acknowledgements}

Figures 1A (right) and 2A created with BioRender.com.

\bibliography{bibliography}

\bibliographystyle{ieeetr}

\end{document}